\documentclass[preprint, pdf]{iucr}
\journalcode{J}
\papertype{FA}
	
\usepackage{siunitx}
\usepackage{mhchem}
\usepackage{amsmath}
\usepackage{amsfonts}
\usepackage{amssymb}
\usepackage{graphicx}
\usepackage{xcolor}
\usepackage{float}
\usepackage{rotating}

\begin{document}

\author[a,c]{Andreas Haahr}{Larsen}
\cauthor[b,c]{Martin Cramer}{Pedersen}{mcpe@nbi.ku.dk}

\aff[a]{Department of Biochemistry, University of Oxford (Linacre College), United Kingdom}
\aff[b]{Niels Bohr Institute, University of Copenhagen, Denmark}
\aff[c]{These authors contributed equally to the presented work}

\title{Experimental noise in small-angle scattering can be assessed using the Bayesian Indirect Fourier Transformation}
\shorttitle{Assessing experimental noise in small-angle scattering}

\maketitle

\begin{synopsis}
We find that the Bayesian Indirect Fourier Transformation algorithm can accurately estimate the noise level of small-angle scattering data. This can be used to: (i) evaluate whether experimental errors are over- or underestimated; (ii) rescale the experimental error estimates; and (iii) determine what reduced $\chi^2$ to aim for in model refinement. 
\end{synopsis}

\begin{abstract}
Small-angle X-ray and neutron scattering are widely used to investigate soft matter and biophysical systems. The experimental errors are essential when assessing how well a hypothesized model fits the data. Likewise, they are important when weights are assigned to multiple datasets used to refine the same model. Therefore, it is problematic when experimental errors are over- or underestimated. We present a method, using Bayesian Indirect Fourier Transformation for small-angle scattering data, to assess whether or not a given small-angle scattering dataset has over- or underestimated experimental errors. The method is effective on both simulated and experimental data, and can be used to assess and rescale the errors accordingly. Even if the estimated experimental errors are appropriate, it is ambiguous whether or not a model fits sufficiently well, as the ``true'' reduced $\chi^2$ of the data is not necessarily unity. This is particularly relevant for approaches where overfitting is an inherent challenge, such as reweighting of a simulated molecular dynamics trajectory against a small-angle scattering data or ab initio modelling. Using the outlined method, we show that one can determine what reduced $\chi^2$ to aim for when fitting a model against small-angle scattering data. The method is easily accessible via a web interface.
\end{abstract}

\section{Introduction}

Small-angle X-ray and neutron scattering (SAXS and SANS) are valuable tools for obtaining structural information in the nanometer-regime for a range of materials, from inorganic nanoparticles, over polymer gels and colloids, to biomolecules in solution. However, some challenges remain. In this paper, we aim to overcome two central challenges:

The first challenge is the issue of over- or underestimated experimental errors. To retrieve structural information from SAXS or SANS data, the data are usually compared to a theoretical model. The assessment of such models does in most cases rely on the estimated experimental errors, with some notable exceptions being the Wald-Wolfowitz runs test and the related CorMap test~\cite{Franke2015}. 

Experimental error estimates are also important when combining a SAXS or SANS dataset with others sources of experimental data, to assess which weights should be assigned to each type of data. This could be e.g. SAXS and SANS~\cite{Larsen2020} or SAXS and nuclear magnetic resonance~\cite{Mertens2017}. Therefore, appropriate experimental errors are essential. Unfortunately, these experimental errors are not always well-determined or well-behaved. This has been demonstrated for the data in the small-angle scattering biological data bank (SASBDB)~\cite{Kikhney2019}.

The second challenge is the question of how tightly a model should be fitted to a dataset. The information content in small-angle scattering data is small compared to the number of degrees of freedom of the three-dimensional structural models, the scientist wishes to refine from the data. To avoid overfitting, prior knowledge or constraints are usually introduced, e.g. from other experiments, from simulations, or from more general assumptions~\cite{Larsen2018}. As an example, molecular dynamics (MD) simulations can be used in combination with SAXS data to refine a model, e.g. by reweighting the frames of the trajectory such that the calculated scattering from the trajectory matches the data. However, such simulated trajectories contain information about the position and movement of thousand of atoms, which far exceeds the $10$ to $20$ free parameters that can be retrieved from a typical SAXS datasets~\cite{Vestergaard2006, Konarev2015}. A good model must therefore alter the prior trajectory as little as possible, but still fit the data sufficiently well. But how well is "sufficiently well"~\cite{Orioli2020}?

One solution is to aim for a reduced $\chi^2$ of one. $\chi^2_r$, which we will discuss in more detail later, is a measure for the goodness of fit and has an expectation value of unity, so aiming for $\chi^2_r=1$ is  appealing and simple. There are, however, several problems with this approach~\cite{Andrae2010_2}; even under the assumption that the experimental errors are appropriate. One problem is that the correct or ``true'' $\chi^2_r$ is not unity for all datasets. Even if we knew the true underlying model and could calculate a corresponding SAXS or SANS model intensity profile accurately, we would not expect a $\chi^2_r$ of exactly unity, but rather one that followed the theoretical $\chi^2_r$ distribution. So, in principle, one should aim for the true, underlying $\chi^2_r$ for that specific dataset when fitting data. The second challenge can therefore be solved by estimating the unknown value of the true $\chi^2_r$ for a given dataset. That value is sometimes referred to as the ``noise level'' of the dataset~\cite{Hansen2000}. 

In the presented study, we demonstrate how this quantity can be estimated with the Bayesian Indirect Fourier Transformation (BIFT) algorithm~\cite{Hansen2000}, which in turn provides an answer to the question: ``when does the model fit the data sufficiently well?'': it does so, when the $\chi^2_r$ of the fit matches the true $\chi^2_r$ of that particular dataset, as estimated by the method presented here. For a specific dataset this may be $2.2$ or $0.4$, i.e. relatively far from the textbook-mandated target value of $1.0$. This is particularly relevant for data with few datapoints, and hence few degrees of freedom, resulting in a wide distribution for $\chi^2_r$. The method is useful for combining SAXS or SANS with MD simulations, and when doing ab initio modeling~\cite{Franke2009} or other types of approaches that involves models with many degrees of freedom. 

The method is based on the fundamental idea of using a low-biased method to estimate experimental errors \cite{Hastie2009}, and thus relates to basic ideas used in, e.g., machine learning \cite{Belkin2019}.

\section{Methods and Software}

\subsection{Small-angle scattering}

The central technique investigated in this study is small-angle scattering, which obtains structural information on a sample as follows. A sample is irradiated by an intense, monochromatic, well-collimated beam of X-rays or neutrons. The scattered radiation is recorded on a position-sensitive detector. For samples without a preferred orientation such as e.g. molecules in solution, the detected pattern exhibits rotational symmetry, and the data can be azimuthally averaged into a one-dimenisional set of data.

We record the scattered intensity, $I$, as a function of the momentum transfer of the incoming radiation, $q$, which is given by $q = 4\pi\sin(\theta)/\lambda$, where $\lambda$ is the nominal wavelength of X-rays or neutrons, and $2\theta$ is the scattering angle. Additionally, data reduction software assign an estimate of the experimental error, $\sigma$, on the intensity recorded for each value of $q$. This error is based on counting statistiscs and error propagation \cite{Svergun1994}. Summing up, a datapoint in a small-angle scattering consists of the triplet $\{q, I, \sigma\}$. 

In some cases, the instrument software provides a fourth column in data, expressing the resolution effects of the instument  \cite{Pedersen1990}, which relates to pixel size, collimation and $\lambda$. This is relevant for SANS data, and SAXS datapoints close the the beam stop, but can in most cases be neglegted for synchrotron SAXS data. This is currently not implemented in BayesApp. 

\subsection{Virtual ray-tracing experiments}

For this study, we simulated thousands of SAXS datasets using a simple, virtual SAXS instrument with dimensions based on the beam line BM29 \cite{Pernot2013} at ESRF in Grenoble. The simulations were done using the X-ray instrument simulation software package McXtrace~\cite{Knudsen2013} using methods outlined in the literature~\cite{Kynde2014, Pedersen2014}.

In the simulations, we assumed a nominal X-ray wavelength of \SI{1}{\angstrom}, a collimation length of \SI{11.1}{m}, \SI{0.7}{mm} pinholes, a beamport-to-sample distance of \SI{0.1}{m}, a sample-to-detector distance of \SI{2.43}{m}, and beam stop covering the incoming beam. The signal was recorded on a two-dimensional position-sensitive detector and azimuthally averaged for a final range in scattering momentum transfer of approximately \SI{0.003}{\per\angstrom} to \SI{0.47}{\per\angstrom}.

We conducted virtual experiments for three distinct samples in solution: the proteins lysozyme and bovine serum albumine (BSA) described by their respective crystal structures (i.e. the entries 2LYZ and 4F5S in the protein data bank~\cite{Berman2000}) as well as a n-dodecyl-$\beta$-D-maltoside (DDM) detergent micelle at \SI{24}{\degreeCelsius} described by the mathematical model implemented in WillItFit~\cite{Pedersen2013} with dimensions taken from experimental reports in the literature~\cite{Oliver2013}. 

To investigate systematic errors, we simulated samples of lysozyme with a fraction of aggregates or with  concentration effects. The aggregates were described as mass fractals with dimensionality of $2$~\cite{Larsen2020} and $100$ lysozyme particles per aggregate, $N$, and $R_g = N^{1/3}R_{g,\mathrm{Lys}}$, where $R_{g,\mathrm{Lys}}$ is the radius of gyration for monomeric lysozyme. The fraction of aggregated lysozyme in the sample was varied from $5\%$ to $80\%$. The concentration effects were described via a hard-sphere structure factor~\cite{Percus1958, Kinning1984}, where the volume fraction was varied from $0.05$ to $0.3$.

The scattering profiles for the proteins were calculated using the software package CRYSOL~\cite{Svergun1995}, whereas scattering from the solution of micelles was computed using geometric form factor amplitudes~\cite{Pedersen1997} as implemented in WillItFit~\cite{Pedersen2013}.

\begin{figure}
	\centering	

	\includegraphics[width=\textwidth]{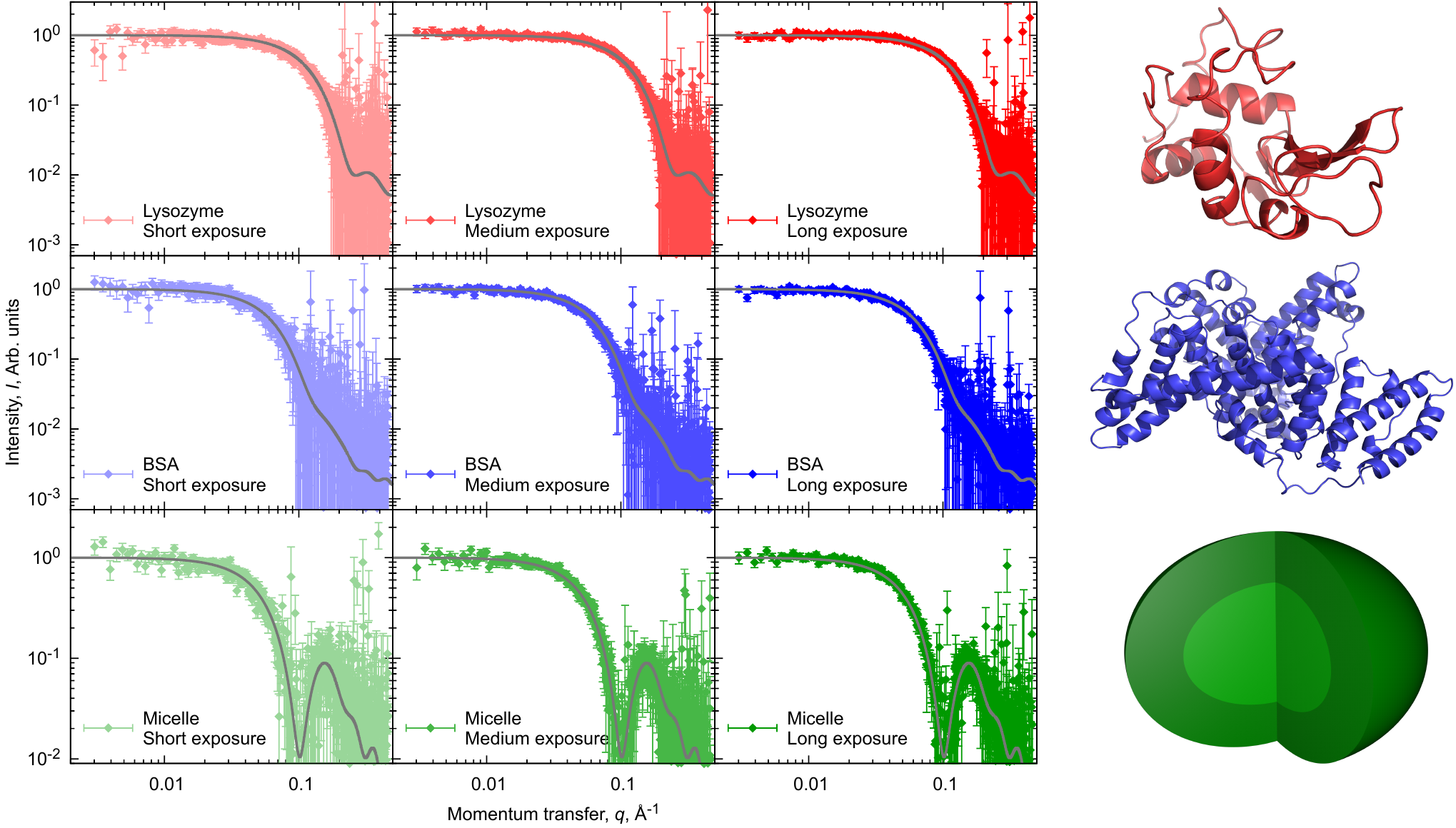}

	\caption{Examples of the simulated datasets for each of our three systems: lysozyme (top), BSA (middle), and DDM micelles (bottom). As shown, we simulate three different exposure times for each of these systems. On the right, the structural models from which the data are simulated are rendered. A quarter of the oblate DDM micelle has been removed to reveal the interior core-shell structure.}

	\label{Figure:SimulatedData}
\end{figure}

Our virtual experiments represent three different exposure times: \emph{short}, \emph{medium}, and \emph{long}, the durations of which differ by a factor of ten. Examples of simulated data are shown in Figure~\ref{Figure:SimulatedData}; and all source code for these components are available through online repositories\footnote{\texttt{https://github.com/McStasMcXtrace/McCode}}. The errors of the simulated data follow the distributions described in the literature \cite{Sedlak2017}. 

\subsection{Bayesian Indirect Fourier Transform and BayesApp}

Since its introduction in the seventies~\cite{Glatter1977}, the Indirect Fourier Transform has been a staple in the preliminary analysis of solution small-angle scattering data by producing pair distance distributions of recorded data for an initial glance at the sample's structure. Popular implementations of the algorithm rely on Bayesian statistics and optimization for unbiased estimates of the parameters needed for the transformation.

Among these are the BIFT algorthm~\cite{Hansen2000} with its associated web-based implementation, BayesApp~\cite{Hansen2012, Hansen2014}. The methodology is readily extended to structural models~\cite{Larsen2018}. A GenApp-based~\cite{Brookes2017} implementation of BayesApp is available online\footnote{\texttt{https://somo.chem.utk.edu/bayesapp}}. The source code can be found in the associated repository\footnote{\texttt{https://github.com/Niels-Bohr-Institute-XNS-StructBiophys/BayesApp}}.

The central objective of the algorithm is to estimate the pair distance distribution, $p(r)$ of the sample, from the measured intensity. We remind the reader that a small-angle scattering intensity profile, $I(q)$, is related to $p(r)$ by \cite{Book-GlatterKratky1982}:
\begin{align}
	p(r) = \frac{1}{2\pi ^2 n} \int_0^\infty \mathrm{d}q \, (qr)^2 I(q) \frac{\sin(qr)}{qr},
	\label{Eqn:pr}
\end{align}
where $n$ is the number density of the given particle in the sample. In an experiment, intensity is measured only in a limited range, and in some discrete points with associated errors, so the integral can not be to evaluated directly. Instead, the BIFT algorithm estimates $p(r)$ in an indirect matter.

First, from an initial guess of $p(r)$, the inverted version of Equation~\eqref{Eqn:pr} is used to calculate the intensity:
\begin{align}
	I(q) = 4 \pi n\int_0^\infty \mathrm{d}r \, p(r) \frac{\sin(qr)}{qr}.
	\label{Eqn:Intensity}
\end{align}
In practice, the integral can be truncated at the largest distance between two scatterers, $D_\text{max}$. $p(r)$ is usually represented via an expansion on a suitable set of basis functions, $\phi_i(r)$, such as e.g. cubic splines or cardinal sine functions:
\begin{align}
	p(r) = \sum_i c_i\phi_i(r).
	\label{Eqn:ci}
\end{align}
The coefficients of the basis functions, $c_i$, are adjusted, until the intensity calculated by Equation~\eqref{Eqn:Intensity} matches the measured intensity. To avoid overfitting, the fitting process is done under a smoothness constraint on $p(r)$~\cite{Glatter1977, Tikhonov1977}. Specifically, the BIFT algorithm minimizes the functional:
\begin{align}
	Q = \chi^2 + \alpha S,
	\label{Eqn:Q}
\end{align}
where $\chi^2$ measures the overlap between a set of $M$ datapoints, ($q_j$, $I_{\text{exp},j}$, $\sigma_j$) and the model intensity from BIFT, $I_\text{mod}(q)$ (Equation \eqref{Eqn:Intensity}), evaluated in $q_j$:
\begin{align}
	\chi^2 = \sum_{j = 1}^M \left(\frac{I_\text{mod}\left(q_j\right) - I_{\text{exp},j}}{\sigma_j}\right)^2,
	\label{Eqn:Chi2}
\end{align}
and $S$ is the prior smoothness constraint:
\begin{align}
	S = \int_0^\infty \mathrm{d}r \, (p''(r))^2
\end{align}
where the $'$ denotes the derivative. 

As before, the upper limit of integral can be truncated at $D_\text{max}$. The hyperparameter $\alpha$ in Equation~\eqref{Eqn:Q} weighs the two contributions; finding the optimal value of $\alpha$ is part of the BIFT algorithm objective \cite{Hansen2000}. Note that alongside $\alpha$, $D_\text{max}$ is also an estimated hyperparameter in this approach.

BIFT also provides an estimate for the number of good parameters, $N_g$, in the dataset. I.e. how many degrees of freedom that were used to fit the data \cite{Vestergaard2006}. From the fitting process, one obtains a Hessian, consisting of the matrix elements $B_{ij} = \frac{\partial^2 \chi^2}{\partial c_i \partial c_j}$, where $c_i$ are the coefficients from Equation $\eqref{Eqn:ci}$. $N_g$ is given via $\alpha$ and the eigenvalues, $\lambda_i$, of the Hessian:
\begin{align}
    N_g = \sum_i^{N_b}\frac{\lambda_i}{\alpha + \lambda_i},
\end{align}
where $N_b$ is the number of basis functions used to represent $p(r)$. 

\subsection{Reduced $\chi^2$}

Of particular importance to this study is the notion of the reduced $\chi^2$, usually dubbed $\chi^2_r$. Any statistics textbook will teach us to normalize the quantity in Equation~\eqref{Eqn:Chi2} using the degrees of freedom $N_\text{DoF}$:
\begin{align}
	\chi^2_r = \frac{1}{N_\text{DoF}} \sum_{j = 1}^M \left(\frac{I_\text{mod}\left(q_j\right) - I_{\text{exp},j}}{\sigma_j}\right)^2.
	\label{Eqn:ReducedChi2}
\end{align}
The pertinent question for further application of this quantity is: What is $N_\text{DoF}$? For a simple model with $K$ independent parameters, $N_\text{DoF}$ is simply equal to the number datapoints minus the number of fitting parameters, $M - K$. 

\begin{figure}
	\centering

	\includegraphics[width=87mm]{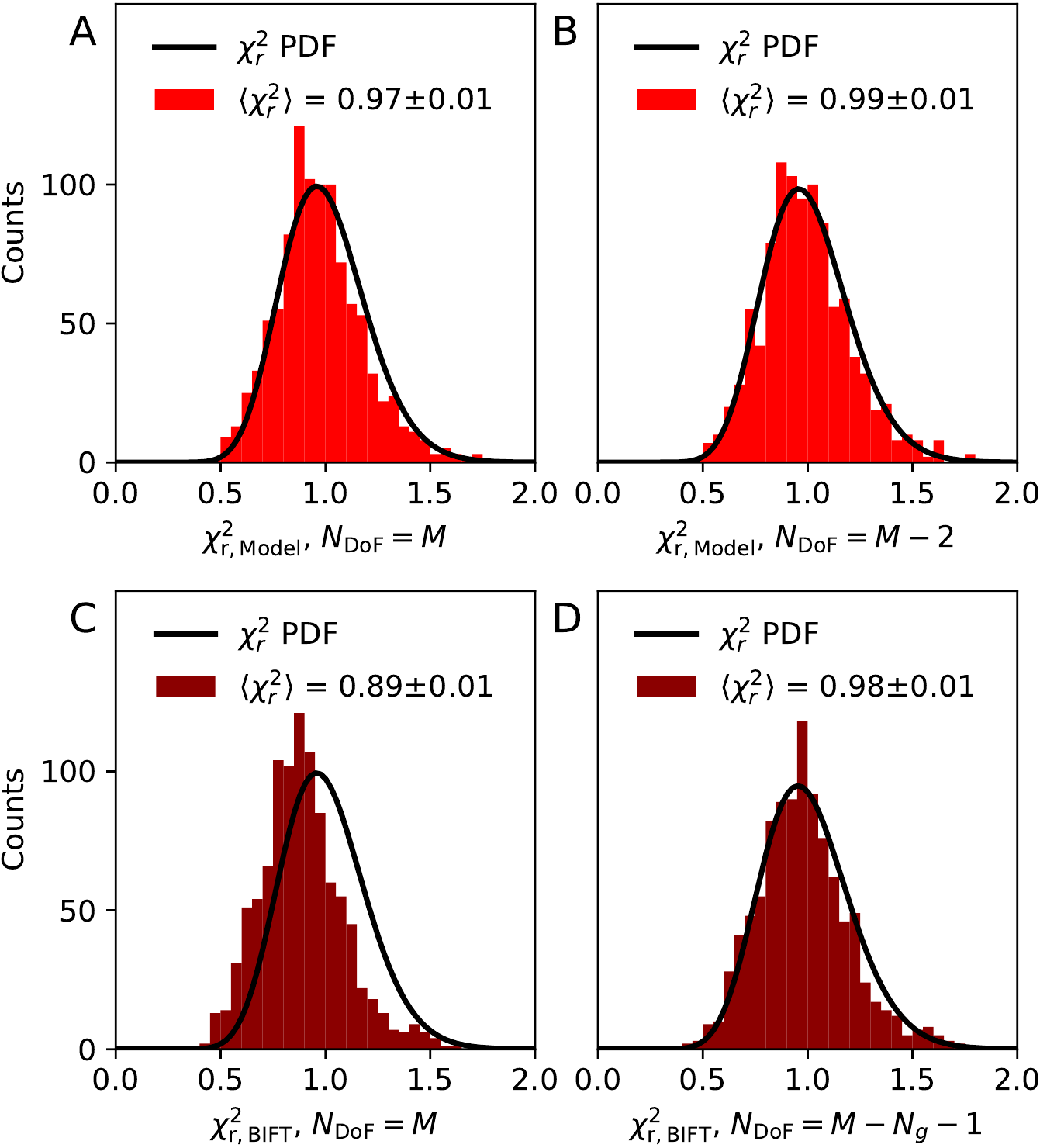}	
	
	\caption{$1000$ SAXS datasets of lysozyme were simulated with long exposure time in McXtrace. Each dataset contained $M = 50$ points; examples of these datasets shown in Figure~\ref{Figure:ExamplesWith50Points}. (A-B) Models were fitted to the data with a scaling factor and a constant background as free parameters with (A) $M$ or (B) $M-2$ as degrees of freedom in the expression for $\chi^2_r$ in Equation~\eqref{Eqn:ReducedChi2}. The average value of $\chi^2_r$, $\langle\chi^2_r\rangle$, is listed in the legend, and the theoretical $\chi^2_r$ probability distribution function (PDF) for the given number of degrees of freedom is shown. (C-D) Histograms of $\chi^2_r$ values from the BIFT fits with (C) $M$ or (D) $M - N_g - 1$ as $N_\text{DoF}$ in $\chi^2_r$.}
	
	\label{Figure:Chi2}
\end{figure}

As an example, this could be a model where a known scattering formfactor is fitted to a dataset with a scaling parameter and a constant background, i.e. with two free parameters. In that case, $N_\text{DoF} = M - 2$ gives the correct distribution of for $\chi^2_r$ (Figure \ref{Figure:Chi2}). In this paper, we will discuss how to estimate $N_\text{DoF}$ for a $p(r)$ distribution based on an expansion on a set of (correlated) basis functions such as BIFT. 

\section{Results}

\subsection{Number of degrees of freedom for a pair distance distribution function}

As described above, the BIFT algorithm performs an analysis resembling that of single value decomposition, to provide the number of good parameters, $N_g$~\cite{Hansen2000, Vestergaard2006}. This quantity is the effective number of free parameters, taking the smoothness prior into account, and is therefore a good estimate for the number of effective paramters in the model. 

BIFT has additional degrees of freedom, as it estimates a constant background, $B$, and a maximum diameter, $D_\mathrm{max}$. If $D_\mathrm{max}$ is too low, the data can not be fitted well, and that is also the case if $B$ is too high. Therefore, we argue that these parameters contribute with one additional degree of freedom to the model. We therefore suggets that $N_\text{DoF} = M - N_g - 1$. This improves of the distribution for $\chi^2_r$ (Figure \ref{Figure:Chi2} and Figure \ref{Figure:Chi2SI}), when compared to simply using $N_\text{DoF} = M$. 

\subsection{Correlation between experimental noise and BIFT}

We investigated whether the noise level of a dataset can be found using the BIFT algorithm. For that purpose, we simulated an extensive amount of virtual data. Unlike experimental data, the noise level can be recovered from simulated data. As we used a model in the form of a calculated SAXS scattering curve to generate the simulated data, we can determine the noise level of the simulated data, $\chi^2_\mathrm{r, Model}$, by fitting the same curve to the simulated data. 

Following this, we ran BIFT for each dataset, which gave our estimated value for the noise level, $\chi^2_\mathrm{r, BIFT}$, and monitored the correlation between $\chi^2_\mathrm{r, Model}$ and $\chi^2_\mathrm{r, BIFT}$.

\begin{figure}
	\centering
	
	\includegraphics[width=\linewidth]{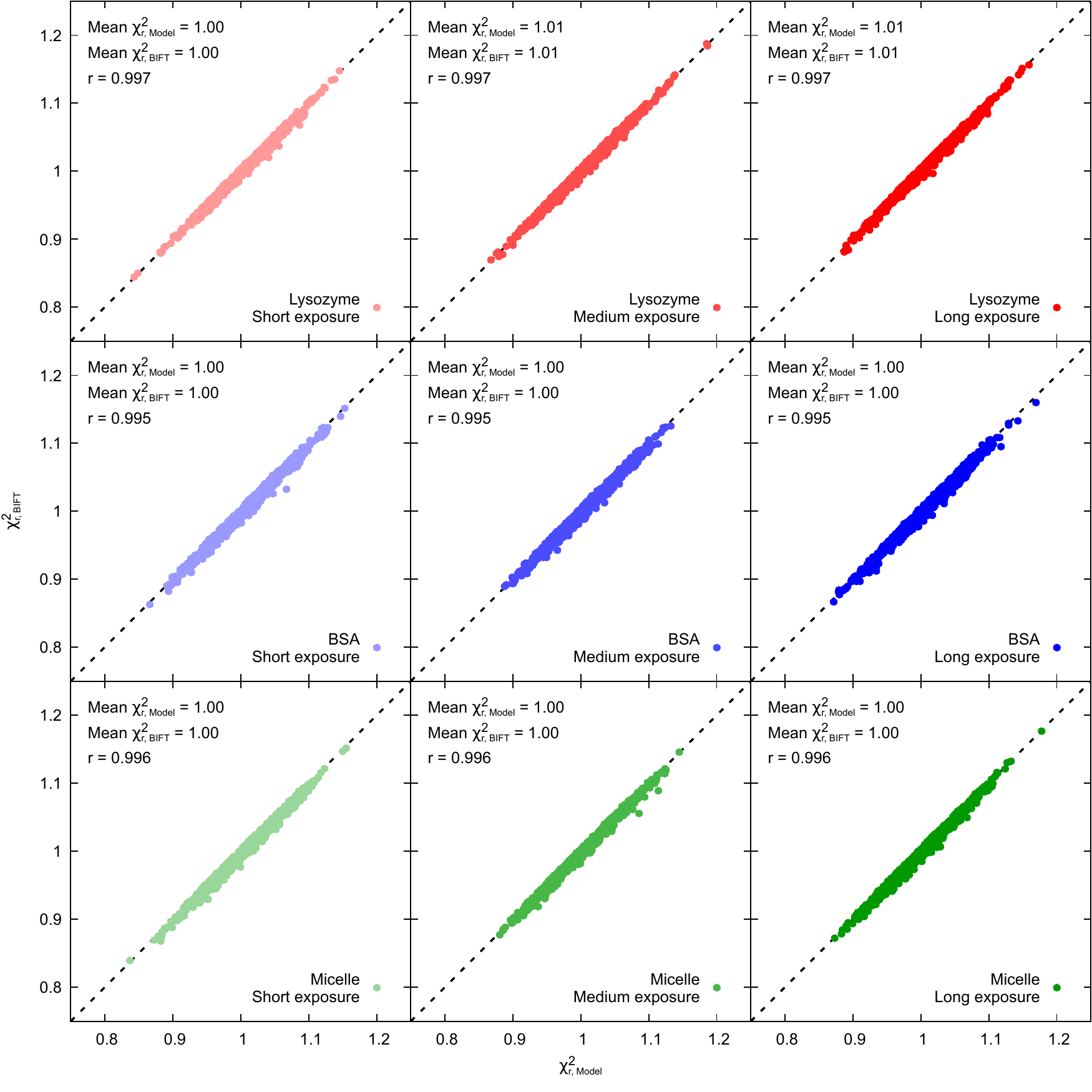}
	
	\caption{Correlation plot of the calculated values og $\chi^2_\text{r, Model}$ and $\chi^2_\text{r, BIFT}$ (using the normalisation introduced in this paper via Figure~\ref{Figure:Chi2}). Pearson Correlation Coefficients, $r$, are listed in each legend of the plots. Examples of these simulated datasets are shown in Figure~\ref{Figure:SimulatedData}.}
	
	\label{Figure:Correlation}
\end{figure}

 The correlation is strong for all tested systems (lysozyme, BSA and a micelle) and all tested exposure times (Figure \ref{Figure:Correlation}), which supports the notion that $\chi^2_\mathrm{r, BIFT}$ is a direct expression of the noise level of the dataset, $\chi^2_\mathrm{r, Model}$. That is a central observation in this study and necessary for being able to further assess and correct over-or underestimated errors.

\subsection{Rescaling errors with BIFT}

Since BIFT can determine the noise level of the data, we propose that it can also be used to identify over- or underestimated errors. If $\chi^2_\text{r, BIFT}$ is much larger than unity it is an indication that errors are underestimated, and if $\chi^2_\text{r, BIFT}$ is much smaller than unity, errors are likely overestimated. In those cases, a better estimate of the errors can be achieved by simply rescaling the experimental errors on the $j$th datapoint: 
\begin{align}
    \sigma_\text{j,Rescaled} = \sigma_\text{j,Recorded} \sqrt{\chi^2_\text{r, BIFT}},
    \label{Eqn:sigma}
\end{align}
where $\sigma_\text{j,Recorded}$ is the over- or underestimated experimental error on the $j$th datapoint. We note that the BIFT algorithm should be run on each setting in a SANS dataset independently and these datasets should be rescaled with each their factor.

\begin{figure}
	\centering
	
	\includegraphics[width=87mm]{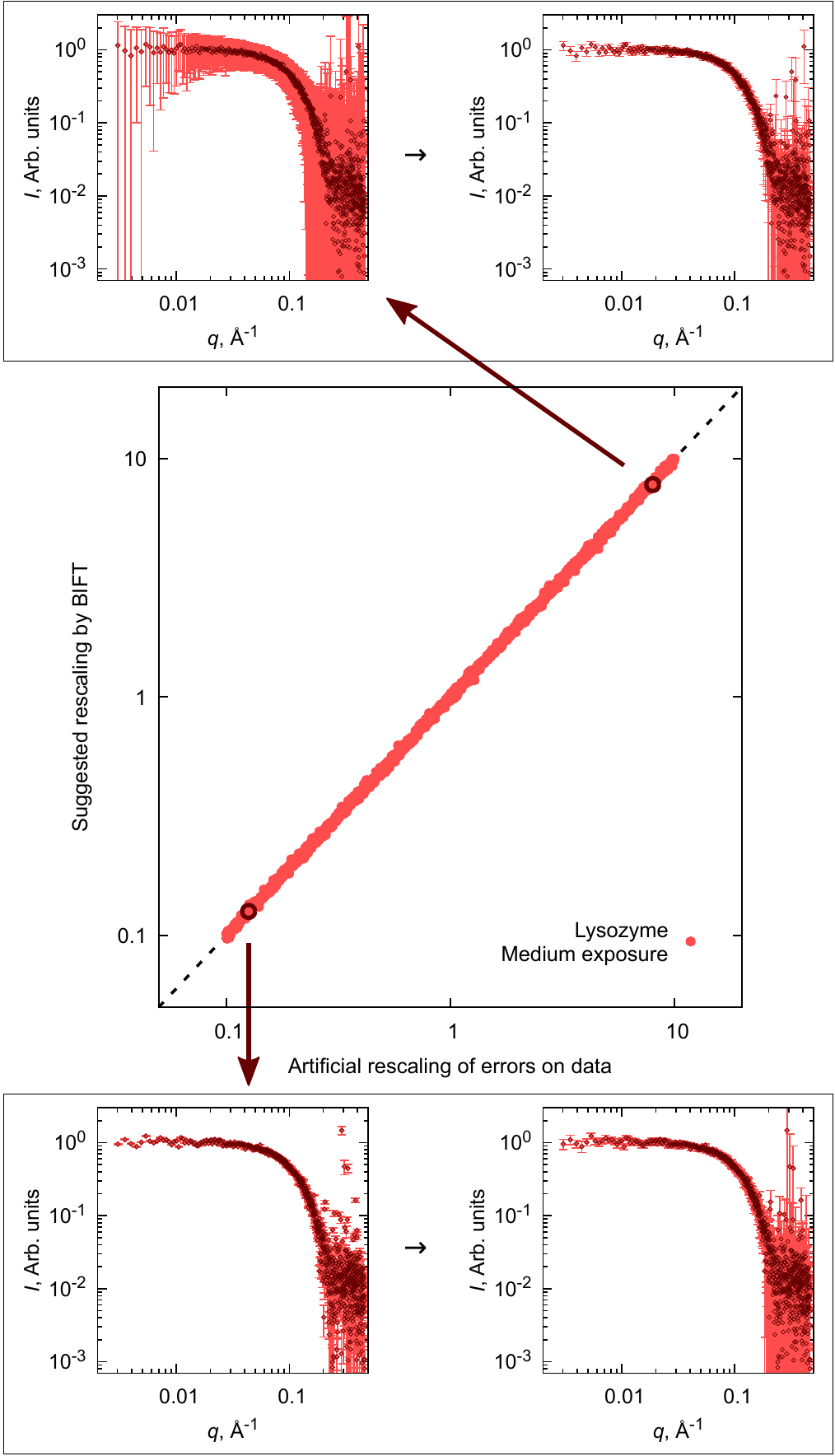}
	
	\caption{The errors on simulated data were multiplied by a factor, effectively changing them to over- or underestimated errors. This factor was estimated using the outlined approach. On top and below, we show examples of datasets rescaled by factors of $8$ and $\frac{1}{8}$, respectively. For completeness, the data from the remaining simulations have been subjected to the same numerical experiment (Figure~\ref{Figure:ErrorSI}.)}
	
	\label{Figure:Error}
\end{figure}

To extend our simulations further, we rescaled the errors of our simulated datasets with factors between $0.1$ and $10$ to mimic over- or underestimated experimental errors. The BIFT algorithm was run on each dataset, and from that a factor to rescale these artificially over- or underestimated errors was obtained using Equation~\eqref{Eqn:sigma}. As shown in Figure~\ref{Figure:Error}, the artificially introduced rescaling factors were accurately recovered by BIFT, demonstrating that BIFT can identify over- or underestimated errors. 

\subsection{Experimental example: SAXS and SANS data fitted with a nanodisc model}

To illustrate the method on real experimental data, we used small-angle scattering data measured on a sample of nanodiscs with the phospholipid DLPC (1,2-dilauroyl-sn-glycero-3-phosphocholine) and the membrane scaffolding protein MSP1D1~\cite{Bayburt2002}. The structural model has been described previously~\cite{Skar-Gislinge2010, Skar-Gislinge2011}. The data consist of five datasets: one SAXS dataset, two SANS datasets measured in $100\%$ \ce{D_2O}-based buffer (high-$q$ and low-$q$ setting), and two SANS datasets measured in $42\%$ \ce{D_2O}-based buffer, where the protein rim of the nanodisc is matched out.

All datasets were Fourier transformed using the BIFT algorithm, which provides a $\chi^2_r$ for each dataset (Table \ref{Table:p-values}). The resulting probabilities show that the $\chi^2_r$ values are unlikely given correctly estimated errors. The SAXS data have underestimated errors ($\chi^2_r$ significantly larger than $1$), whereas all the SANS data have overestimated errors ($\chi^2_r$ significantly smaller than $1$). 

\begin{figure}
	\centering
	
	\includegraphics[width=87mm]{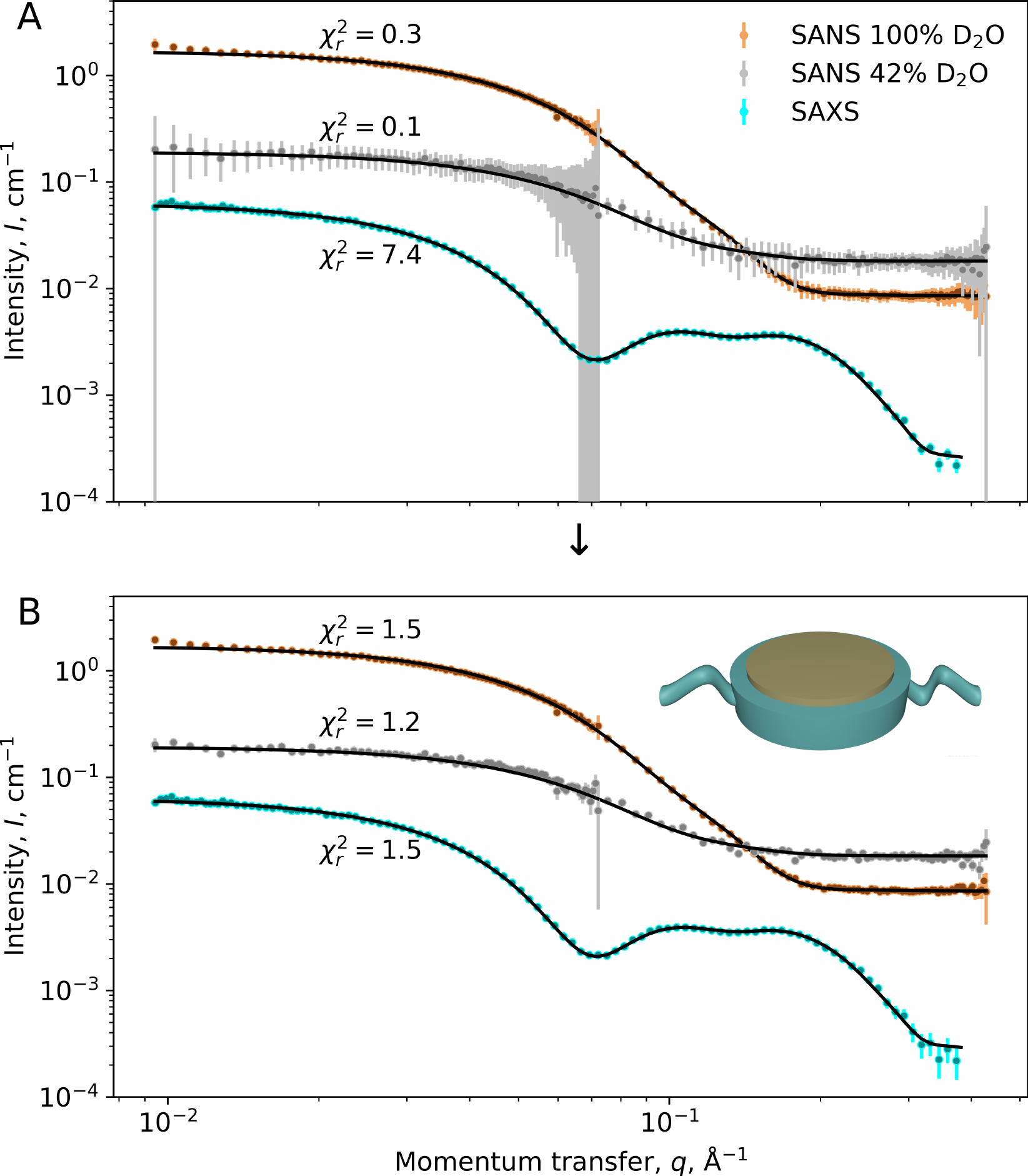}
	
	\caption{SAXS and SANS data for a sample of phospholipid bilayer nanodiscs. A geometrical nanodic model refined from the data is shown as inset, and the model fit is shown in black, with resulting $\chi^2_r$ values given on the plots. (A) Data and fits before rescaling the errors with BIFT. (B) Data and fits after rescaling. Inset shows the nanodisc model.}
	
	\label{Figure:Example}
\end{figure}

The data can simultaneously be fitted with a nanodisc model (Figure~\ref{Figure:Example}A). Despite good fits, the resulting $\chi^2_r$ values from the model fits are, respectively, very small (SANS) or very high (SAXS). We rescaled the errors using the protocol presented here, and after the rescaling, the $\chi^2_r$ values corresponded well with the visual assessment, namely that of a good fit (Figure \ref{Figure:Example}B). The calculated errors on the refined parameters were also more reasonable. Before rescaling the errors, the estimated errors on the refined parameters were underestimated from the fit to SAXS data, and overestimated from the fit to SANS data (Table~\ref{Table:ExampleParams}). 

\begin{table}
	\centering
	
	\begin{tabular}{llll}
		\hline
		Data & BIFT, $\chi^2_r$ & Probability & Rescaling factor\\
		\hline
		SAXS                              & $5.0$  & $\sim 10^{-49}$ & 2.2 \\
		SANS, $100\%$ \ce{D_2O}, low-$q$  & $0.13$ & $\sim 10^{-16}$ & 0.36 \\
		SANS, $100\%$ \ce{D_2O}, high-$q$ & $0.18$ & $\sim 10^{-18}$ & 0.42\\
		SANS, $42\%$ \ce{D_2O}, low-$q$   & $0.02$ & $\sim 10^{-34}$ & 0.14\\
		SANS, $42\%$ \ce{D_2O}, high-$q$  & $0.05$ & $\sim 10^{-51}$ & 0.22 \\
		\hline
	\end{tabular}
		
	\caption{$\chi^2_r$ values from BIFT for each dataset, along with the probabilities for getting these values given that errors are correct (see definition in Supporting Information), and rescaling factors (Equation $\eqref{Eqn:sigma}$). There are two setting for each contrast of the SANS data as seen in Figure~\ref{Figure:Example}.}
	\label{Table:p-values}
\end{table} 

When fitting all dataset simultaneously, such a rescaling will assign a more appropriate weight to each dataset. In this case, more weight is given to the SANS data after rescaling the errors, and, accordingly, less to the SAXS data. For these specific data and the model, the SAXS data is, however, dominating the refinement process, also after rescaling (Table~\ref{Table:ExampleParams}), but for other data and models that is not the case (see e.g. \cite{Heller2020}). Another experimental example can be found in an early use of the method~\cite{Martin2019} with SANS data on the transmembrane protein holo-translocon.

\subsection{Constant or $q$-dependent rescaling}

The algorithm can apply either constant or $q$-dependent rescaling. Constant rescaling is simpler, but assumes that the experimental errors are over/underestimated evenly across the entire $q$-range. This is not always the case, as seen by the residuals for fitting the presented SAXS data (Fig. \ref{Figure:Residuals}). Therefore, we implemented optional $q$-dependent rescaling of errors to BayesApp. If opted for, the data are divided into bins, each containing one Shannon channel~\cite{Shannon1948}. Each Shannon-bin must contain a minimum of 10 points, so If a bin contains fewer than 10 points, it is iteratively merged with the next bin. The $\chi^2_r$ is calculated for each bin, and the errors are rescaled bin-by-bin using a $q$-dependent version of Equation~\eqref{Eqn:sigma}:
\begin{align}
\sigma_\text{j,k,Rescaled} = \sigma_\text{j,k,Recorded} \sqrt{\chi^2_\text{r,BIFT,k}},
\label{Eqn:sigma_bin}
\end{align}
where $\sigma_\text{j,k,Recorded}$ is rhe $j$th point in the $k$th bin, and $\chi^2_\text{r,BIFT,k}$ is the value of $\chi^2_r$ calculated for that bin.

If the errors are over/underestimated in a $q$-dependent manner, this method leads to better rescaling of the errors, as judged by the residuals for fit to the example data after rescaling (Figure \ref{Figure:Residuals}).

\subsection{Data with systematic errors}

The tests and results shown in previous sections demonstrate the applicability of the pipeline in somewhat idealized circumstances; in particular for the sample which is assumed to be a monodisperse solution of the system in question. As part of this study, we additionally tested the BIFT algorithm's ability to ascertain noise levels in data exhibiting systematic deviations from this idealized scenario.

We deployed the framework in two settings: one in which a sample of lysozyme was assumed to aggregate according to the model as discussed earlier, and one in which the sample of lysozyme was assumed to be at a concentration where the effects of a hard-sphere structure factor were evident. In Figures~\ref{Figure:AggrData} and~\ref{Figure:HSData}, we show examples of simulated sets of data and their $p(r)$ distributions under these conditions.

We tested degrees of aggregation ranging from $5\%$ to $80\%$, and volume fractions for our hard-sphere structure factor between $0.05$ and $0.3$ (Figure~\ref{Figure:SystematicCorrelation}). For the parameters values investigated here, we observe how the BIFT algorithm is robust against these disruptions to the idealized conditions, i.e. how the noise level of the simulated data is recovered by considering the $\chi^2$ from the BIFT refinement.

\section{Discussion}

In this paper, we outlined a proposed approach to two central challenges in the interpretation of SAXS and SANS data. The first challenge is the issue of over- or underestimated experimental errors. The second was what value of $\chi^2_r$ to aim for when analysing SAXS or SANS data.

The BIFT algorithm provides a valuable tool for addressing these challenges and can assess experimental noise and obtain a better estimate of the experimental errors. Here, we discuss important limitations of the method and provide some general guidelines for which type of data the method can (and cannot) be applied for. Moreover, we discuss when a rescaling of errors is appropriate and whether to apply constant or $q$-dependent rescaling. 

\subsection{Applicability to data from various systems}

BIFT and its implementation in BayesApp have been designed for soluble systems, e.g. proteins, micelles, etc.~\cite{Hansen2000,Vestergaard2006}. Here, we have shown that the algorithm is also applicaple for data with systematic errors in the form of a fraction of aggregates or in the form of concentration effects.

In this context, we showed that the method is robust against two models for systematic deviations from simple monodispersity. However, we remind the reader that the BIFT algorithm is fundamentally based on the premise that the $p(r)$ distribution is finite and smooth. In the case of extremely ordered samples such as e.g. crystalline samples or samples with sizes above those resolvable by small-angle scattering data, this assumption might be invalid. In these cases, other regularization functionals must be employed and tested. So the method is, at present, limited to soluble samples within the fields of structural biology or soft matter science. 

\subsection{Alternative to rescaling of errors}

Even if data has over- or underestimated errors, a rescaling of the errors may not be necessary for subsequent analysis. Instead, one can aim for a $\chi^2_r$ that is equal to $\chi^2_{r,\mathrm{BIFT}}$ when refining a model against the data. This is a safe choice and does not alter the original data. Moreover, this is also helpful if data has correctly estimated errors, as the BIFT algorithm assesses the noise level in data, i.e. find the correct $\chi^2_r$ to aim, which may differ significancly from unity (Figure \ref{Figure:Correlation}), especially for datasets with few datapoints (Figure \ref{Figure:ExamplesWith50Points}). 

However, the approach of aiming for $\chi^2_{r,\mathrm{BIFT}}$ lacks the convenient and familiar property of $\chi^2_r$ being close to unity for a good fit. Moreover, it is not applicable when combining several datasets, as poorly estimated errors on one dataset would give an incorrect weight between the different types of data included in the analysis. In that case, rescaling of errors is preferable.

\subsection{Criteria for rescaling}

As discussed in the introduction, $\chi^2_r$ may be different from unity even though errors are correct. That is: one has to assess whether the $\chi^2_r$ is so unlikely that it is reasonable to believe that experimental errors are over- or underestimated. For that purpose, one can use the probability of getting a certain value of $\chi^2_\mathrm{r, BIFT}$ given that the errors are correct. This is now given as output in BayasApp. As a criteria for rescaling, we recommend that if the probability for the $\chi^2_\mathrm{r, BIFT}$ value is below $0.003$ (corresponding to three $\sigma$, see Supporting Information), the errors should be rescaled to achieve a better estimate. However, we always recommend thorough inspeciton of the residuals (see, e.g., Figure \ref{Figure:Residuals}), that may reveal, e.g., systematic errors or $q$-dependent variation in the residuals. 

\subsection{Constant or $q$-dependent rescaling}

Adjusting the experimental errors using Equation~\eqref{Eqn:sigma} assumes that uniformly rescaling these across the full $q$-range is appropriate. This was not the case for the example SAXS data (Figure \ref{Figure:Residuals}), where the errors are mostly underestimated in the mid-$q$ range. Therefore, we implemented an option for $q$-dependent rescaling in BayesApp, which gave better results for that dataset (Figure \ref{Figure:Residuals}). However, it is important to note that $q$-dependent rescaling will change the minimum of a fitted model, as the relative weight given to the datapoints is altered. Therefore, this method should be used with care and we recommend first trying a constant rescaling, and only after careful inspection of the residuals, and if necessary, apply $q$-dependent rescaling. 

\subsection{Distribution of $\chi^2_r$ after rescaling}

If the experimental errors are rescaled, and a model subsequently fitted to the data, the resulting $\chi^2_{r,\mathrm{Rescaled}}$ will not follow a $\chi^2_r$ distribution. The BIFT algorithm accurately determines the noise level of data (Figure~\ref{Figure:Correlation}), so $\chi^2_{r,\mathrm{Rescaled}}$ will follow a very narrow distribution around unity.

%
 
\section{Conclusion}

The BIFT algorithm makes for an attractive component in a pipeline for assessing noise in small-angle scattering data as the algorithm is fast, automated, and performs well using default settings for generic small-angle scattering data. The approach is applicable for SAXS and SANS data.

Our findings and associated recommendations can be summed up in the following points:
\begin{itemize}
    \item{The target $\chi^2$ for model refinement should be that of the BIFT algorithm.}
    \item{Alternatively, error bars can be rescaled using the BIFT algorithm. In that case, the target $\chi^2_r$ is one, This is particularly useful for simultaneous refinement of several datasets, e.g. SAXS and SANS, in order to obtain the correct weighting between all data. Note that the emerging  distribution of $\chi^2_r$ after rescaling is not $\chi^2$-distributed.}
\end{itemize}

We would like to stress that the method presented here does not replace rigourious data acquisition and data reduction, leading to correctly estimated experimental errors. This is preferable and treated elsewhere \cite{Svergun1994,Pauw2017}. Instead, the current method is useful for (i) detecting if the errors are over- or underestimated, and for (ii) using the data in the best possible way despite having over- or underestimated errors, when these can not be corrected by redoing the data reduction.

The method is accessible and effective. The interface for Bayesapp\footnote{\texttt{https://somo.chem.utk.edu/bayesapp}} has been updated to support the ideas presented here, where $\chi^2_\text{BIFT}$ will be given along with the probability of that value given that experimental errors are correct. Additionally, a dataset with rescaled experimental errors is produced.

\ack{
The authors would like to thank Steen Laugesen Hansen, who conceptualized the idea and wrote the original implementation of the BIFT algorithm. The authors would also like to thank Emre Brookes for great support in updating the web interface for Bayesapp available through Genapp. Furthermore, the authors thank the Carlsberg Foundation for funding AHL (grant CF19-0288), as well as the Lundbeck foundation (Brainstruc grant R155-2015-2666) and the Novo Nordisk Foundation (Synergy grant NNF15OC0016670) for funding to MCP.}

\bibliographystyle{iucr.bst}
\bibliography{reference.bib}

\clearpage

\section{Supporting Information}

\renewcommand\theequation{SI.\arabic{equation}}
\setcounter{equation}{0}

\renewcommand\thefigure{SI.\arabic{figure}}
\setcounter{figure}{0}

\renewcommand\thetable{SI.\arabic{table}}
\setcounter{table}{0}

\renewcommand\thesubsection{SI.\arabic{subsection}}
\setcounter{subsection}{0}

\subsection{Examples of simulated data with $M = 50$ points}

\begin{figure}
	\centering
	
	\includegraphics[width=\linewidth]{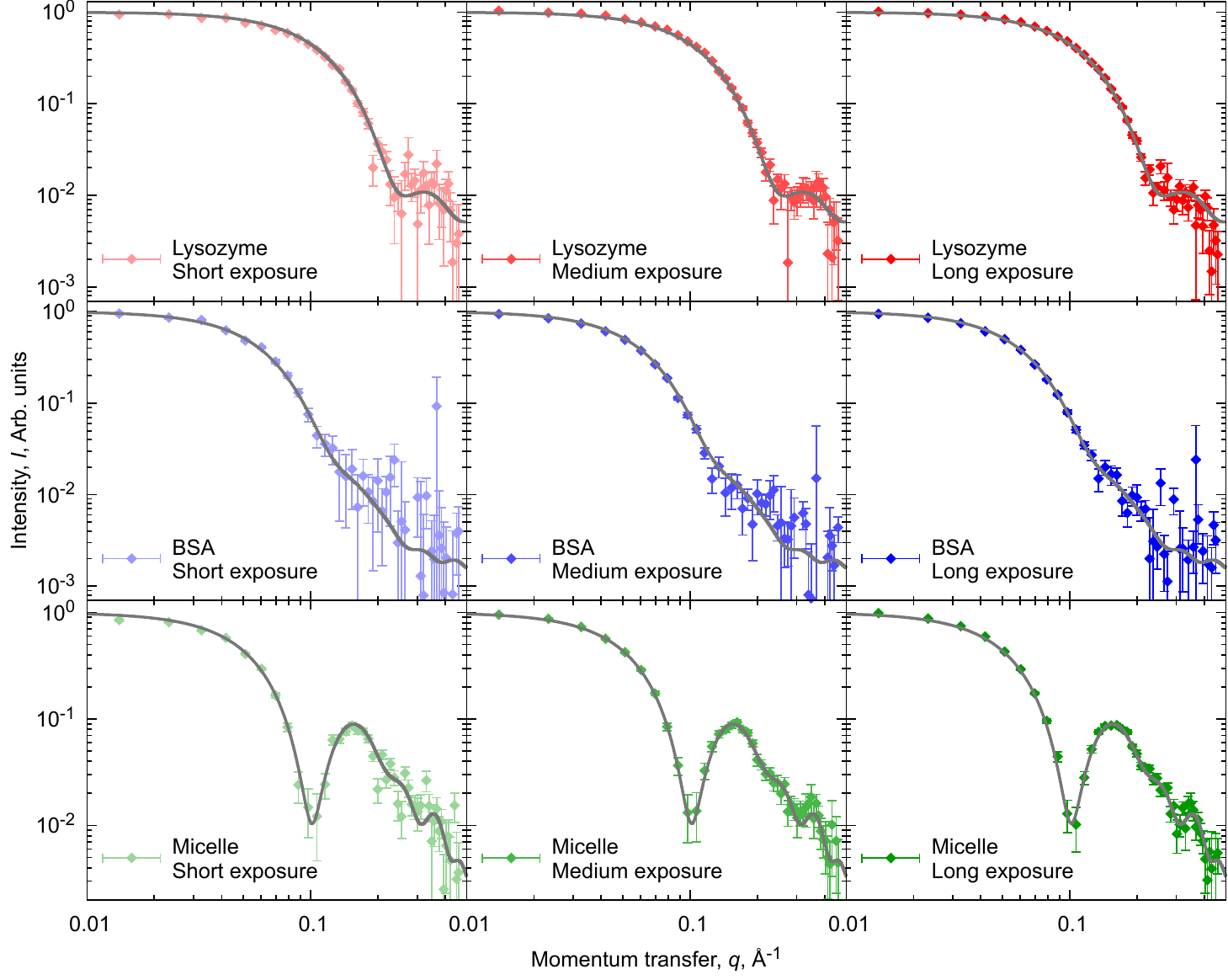}
	
	\caption{Examples of simulated data with $50$ datapoints.}
	
	\label{Figure:ExamplesWith50Points}
\end{figure}

Figure~\ref{Figure:ExamplesWith50Points} shows examples of simulated datasets with a detector binning the data into $50$ bins. The data form the basis for Figures~\ref{Figure:Chi2} and~\ref{Figure:Chi2SI}.

\clearpage

\subsection{Full version of the correlation plots}

\begin{figure}
	\centering
	
	\includegraphics[width=\linewidth]{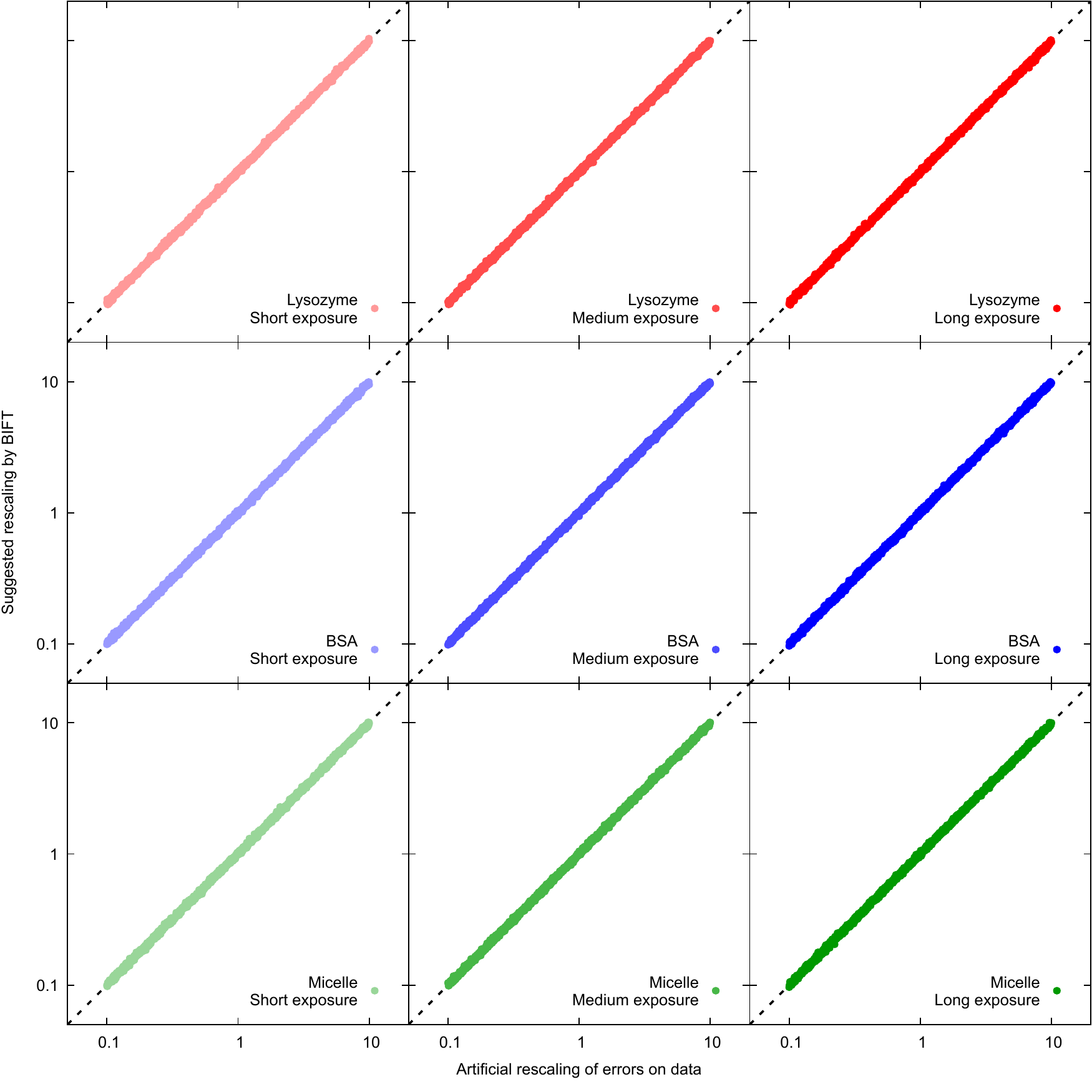}
	
	\caption{Full version of Figure~\ref{Figure:Error}.}
	
	\label{Figure:ErrorSI}
\end{figure}

The full version of the correlation plot in Figure~\ref{Figure:Error} for the versions of our data with rescaled errors are shown in Figure~\ref{Figure:ErrorSI}.

\clearpage

\subsection{Distributions of $\chi^2$s for different choices of $N_\text{DoF}$}

\begin{figure}
	\centering
	
	\includegraphics[width=\linewidth]{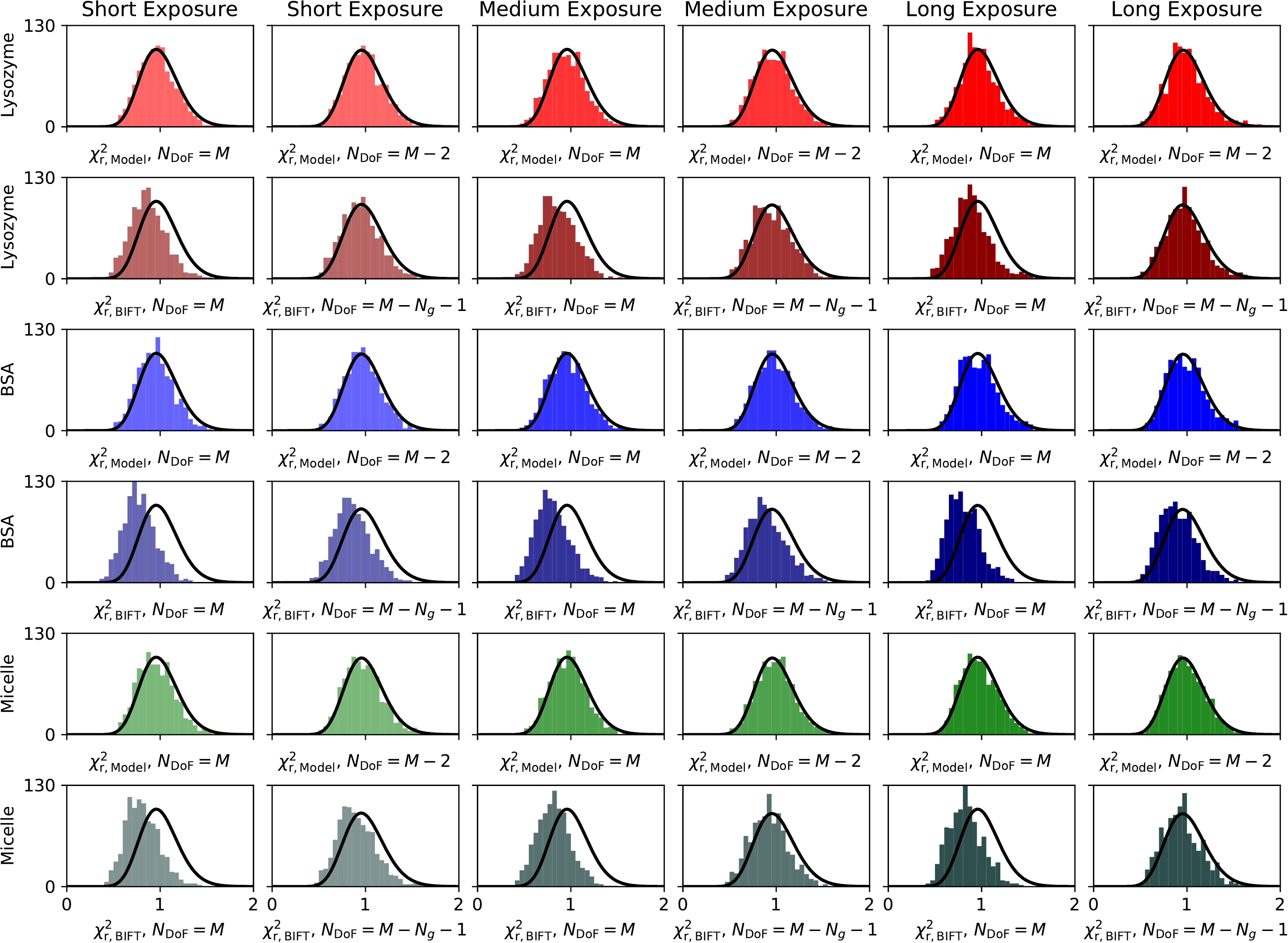}
	
	\caption{Full version of Figure~\ref{Figure:Chi2}. Average values for each histogram given in Table~\ref{Table:Chi2}.}
	
	\label{Figure:Chi2SI}
\end{figure}

All distributions of $\chi^2$ for our datasets with $50$ datapoints for the BIFT algorithm and for the structural model with different choices for $N_\text{DoF}$ are shown in Figure~\ref{Figure:Chi2SI}. Corresponding statistics can be found in Table~\ref{Table:Chi2}.

\begin{table}
	\centering
	
 	\begin{tabular}{llcccc}
 		\hline
 		Model & Exposure time & \multicolumn{4}{c}{Choice of $\chi^2_r$} \\
 		      &               & $\frac{\chi^2_\mathrm{Model}}{M}$ & $\frac{\chi^2_\mathrm{Model}}{M-2}$ &  $\frac{\chi^2_\mathrm{BIFT}}{M}$ & $\frac{\chi^2_\mathrm{BIFT}}{M - N_g - 1}$ \\
 		\hline
 		Lysozyme & Short  & 0.95 & 0.99 & 0.87 & 0.95 \\
 		-        & Medium & 0.93 & 0.97 & 0.86 & 0.95 \\
 		- 	     & Long   & 0.95 & 0.99 & 0.89 & 0.98 \\
 		BSA      & Short  & 0.93 & 0.97 & 0.78 & 0.87 \\
 		-        & Medium & 0.94 & 0.98 & 0.80 & 0.90 \\
 		-        & Long   & 0.95 & 0.99 & 0.81 & 0.90 \\
 		Micelle  & Short  & 0.94 & 0.94 & 0.82 & 0.90 \\
 		-        & Medium & 0.95 & 0.95 & 0.82 & 0.92 \\
 		-        & Long   & 0.95 & 0.95 & 0.83 & 0.94 \\
 		\hline
 		\multicolumn{2}{l}{Mean of average values} & 0.94 & 0.98 & 0.83 & 0.92 \\
 	    \multicolumn{2}{l}{Standard deviation}     & 0.01 & 0.01 & 0.03 & 0.03 \\
 		\hline
 	\end{tabular}
 	
	\caption{Average of the $1000$ $\chi^2_r$ values for each sample/exposure time. Histograms for the same values are displayed in Figure \ref{Figure:Chi2SI}. $M=50$ is the number of datapoints, and $N_g$ is the number of ``good parameters'' from the BIFT algorithm.}
	
	\label{Table:Chi2}
\end{table}

\clearpage

\subsection{Structural parameters for the presented SAS fits}

The parameters refined from the fits in Figure~\ref{Figure:Example} can be found in Table~\ref{Table:ExampleParams}.

\begin{table}
	\centering
	
	\tiny
	
	\begin{tabular}{ l c c c c c c c c c }
		\hline\\
		& \multicolumn{3}{c}{SAXS only} & \multicolumn{3}{c}{SAXS and SANS} & \multicolumn{3}{c}{SANS only} \\
		& \multicolumn{3}{c}{---------------------------------------} & \multicolumn{3}{c}{---------------------------------------} & \multicolumn{3}{c}{---------------------------------------} \\
		& Original         & Rescaled       & Rescaled        & Original       & Rescaled      & Rescaled       & Original      & Rescaled   & Rescaled\\
		&                      & Constant      & q-dep.           &                    & Constant     & q-dep.           &                  & Constant   & q-dep.\\		
		$\chi^2_r$                                       & $1.4$            & $7.2$     & $1.4$     & $1.8$          & $1.3$         & $1.4$  & $0.08$        & $0.94$ & $0.94$ \\	
		& & & & & & & & &\\
		\multicolumn{6}{l}{Structural parameters} & & &\\
		& & & & & & & & & \\
		$\epsilon$                                      & $1.4\pm 0.2$   & $1.4\pm 0.3$    & $1.4\pm0.3$ & $1.4\pm 0.3$   & $1.4\pm 0.5$   & $1.3\pm0.2$ & $1.5 \pm 2.0$     & $1.5\pm 0.5$ & $1.4\pm0.6$\\
		$A$, \SI{}{\angstrom^2}               & $59.4\pm 1.3$  & $59.5\pm 2.9$   & $59.5\pm2.9$ & $59.4\pm 2.1$  & $58.9\pm 2.7$  & $59\pm1$ & $67\pm 222$       & $70\pm 67$ & $70\pm64$ \\
		$N$                                          & $149\pm 12$    & $149\pm 27$     & $149\pm27$ & $152\pm 10$    & $153\pm 5$     & $152\pm4$ & $139\pm 499$      & $143\pm 205$ & $143\pm193$ \\
		$R_{g,tag}$, \SI{}{\angstrom}      & $12\pm 7$      & $12\pm 16$     & $12\pm16$  & $10\pm 11$     & $11\pm 17$     & $9\pm11$ & $43\pm 1015$      & $29\pm 339$ & $30\pm341$ \\
		$V_{MSP}$, \SI{e3}{\angstrom^3} & $26.5\pm 0.6$ & $26.5\pm 1.3$ &  $26.5\pm1.3$ & $26.6\pm 0.8$ & $26.6\pm 1.0$ & $26.7\pm0.6$ & $24\pm 175$ & $19\pm 64$ & $19\pm42$ \\
		$V_{lip}$ \SI{}{\angstrom^3}           & $1006\pm 7$     & $1006\pm 16$   & $1006\pm15$ & $1005\pm 9$     & $1005\pm 12$     & $1004\pm5$ & $1143\pm 3513$     & $1186\pm 1021$ & $1191\pm1187$\\
		& & & & & & & & &\\
		\multicolumn{6}{l}{Contrast-specific parameters} & & &\\
		& & & & & & & & & \\
		$\sigma_{X}$, \SI{}{\angstrom}                                & $4.1\pm 0.4$ & $4.1\pm 0.9$ & $4.1\pm0.9$  & $4.2\pm 0.6$ & $4.3\pm 1.2$ & $4.3\pm0.6$ & --           & -- & --- \\
		$B_{X}$, \SI{e-4}{\per\centi\meter}                     & $2\pm 1$     & $2\pm 3$     & $2\pm3$ & $2\pm 3$     & $2\pm 6$     & $2\pm4$ & --           & -- & --- \\
		$\sigma_{N}$, \SI{}{\angstrom}                                & --           & --           & --- & $4.7\pm 2.9$ & $4.8\pm 1.3$ & $4.9\pm1.1$ & $5.0\pm 7.3$ & $5.1\pm 6.9$ & $5.1\pm6.4$ \\
		$B_{100}$, \SI{e-4}{\per\centi\meter} & --           & --           & --- & $9\pm 4$     & $9\pm 1$     & $85\pm8$ & $9\pm 3$     & $9\pm 1$ &  $85\pm7$\\
		$B_{0}$, \SI{e-4}{\per\centi\meter}  & --           & --           & --- &  $250\pm188$  & $252\pm43$   & $251\pm32$ & $249\pm162$   & $249\pm38$ & $249\pm28$ \\
		\hline
	\end{tabular}
	
	\caption{Parameters refined during the fitting of a phospholipid nanodisc model to data in Figure~\ref{Figure:Example} before rescaling and after rescaling the errors using BIFT. The model is described in the literature~\cite{Skar-Gislinge2010, Skar-Gislinge2011}. The nanodisc model was refined from, respectively, SAXS alone, SAXS and SANS together, or SANS data alone (including SANS samples with $42\%$ \ce{D_2O} and $100\%$ \ce{D_2O} in the buffer). $\epsilon$: axis ratio of bilayer, $A$: area per lipid headgroup, $N$: number of lipids per nanodisc, $R_{g,tag}$: radius of gyration of histidine tag,  $V_{MSP}$: volume of membrane scaffolding protein, $V_{lip}$: volume of phospholipid, DLPC, $\sigma_{X}$: SAXS Roughness, $B_{X}$: SAXS background, $\sigma_{SANS}$: SANS Roughness, $B_{100}$: SANS ($100\%$ \ce{D_2O}) background, $B_{0}$: SANS ($0\%$ \ce{D_2O}) background.}
	
	\label{Table:ExampleParams}
\end{table}

\clearpage

\subsection{Details on the probability of $\chi^2_\mathrm{r, BIFT}$}

\begin{figure}
	\centering
	
	\includegraphics[width=\linewidth]{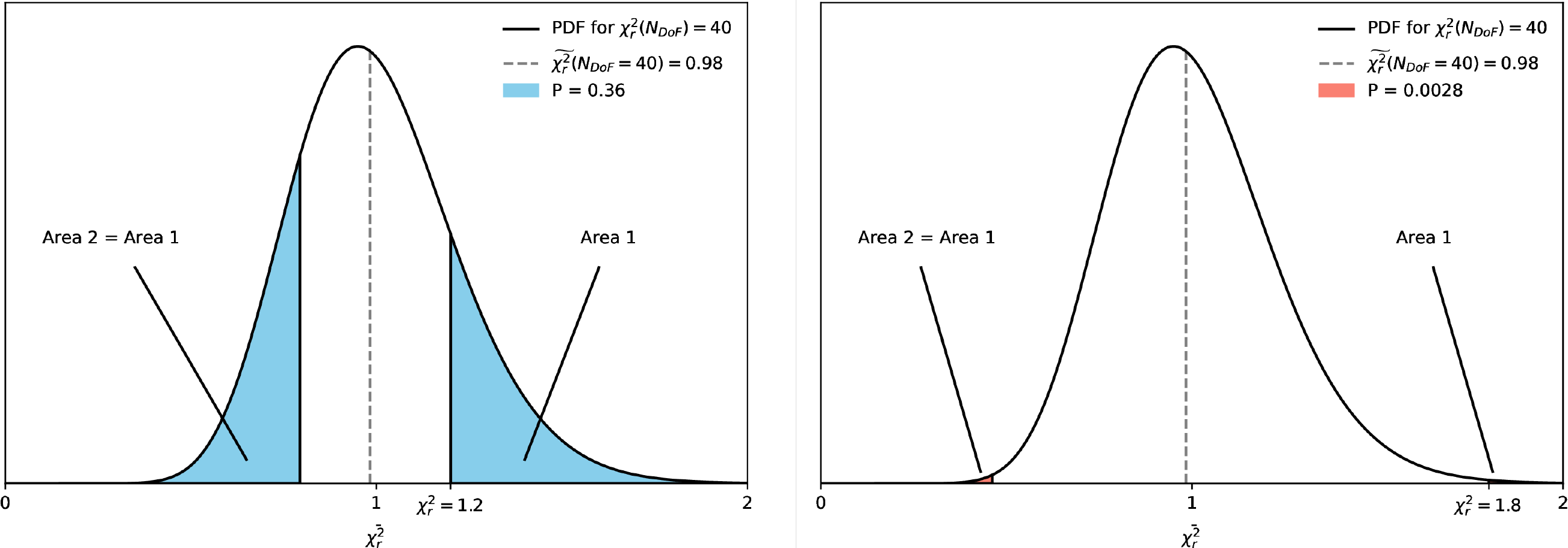}
	
	\caption{Probability of a given $\chi^2_r$ value, illustrated for two different $\chi^2_r$ values and $N_\text{DoF}=40$. The probability equals the area under the graphs for all $\bar{\chi^2_r}\ge\chi^2_r$ plus the same area from the left-side tail. On the left, $\chi^2_r=1.2$ gives $P=0.36$ and is thus far above our suggested significance level of 0.003, so errors are probably correct. On the right, $\chi^2_r=1.8$ and $P=0.0028$, i.e. below the significance level, so the errors are probably underestimated.}
	
	\label{Figure:Prob}
\end{figure}

We use the probability of a given value of $\chi^2_\mathrm{r, BIFT}$ to assess whether the experimental errors in a given dataset are appropriate. This is the probability for getting a particular $\chi^2_{r, \mathrm{BIFT}}$ or any value more extreme, which can be compute as: 
\begin{align}
	 P(\chi_r^2) = \left\{\begin{tabular}{ll} $2\displaystyle\int_0^{\chi_r^2} \, \mathrm{d}\bar{\chi_r^2} \,  p(\bar{\chi_r^2})$ & $\text{for } \chi_r^2 \leq \widetilde{ \chi_r^2}$ \\
	 & \\
	 $2\displaystyle\int_{\chi_r^2}^\infty \, \mathrm{d}\bar{\chi_r^2} \, p(\bar{\chi_r^2})$ & $\text{for } \chi_r^2 > \widetilde{ \chi_r^2}$
	 \end{tabular}\right.
\end{align}
where $p(\bar{\chi_r^2})$ is a probability density that depends on the variable $\bar{\chi_r^2}$. $\widetilde{ \chi_r^2}$ is the median and can be approximated by:
\begin{align}
	\widetilde{ \chi_r^2} \approx \left(1-\frac{2}{9N_\text{DoF}}\right)^3,
\end{align}
where $N_\text{DoF}$ is the degrees of freedom. The median is unity for large $N_\text{DoF}$. $P$ is the two-tailed $P$-value for $\chi^2_\mathrm{r, BIFT}$ given the null hypothesis that the experimental errors in question are appropriate (Figure~\ref{Figure:Prob}). The probability is unity when $\chi^2_r = \widetilde{ \chi_r^2}$, i.e. in most practical cases, when $\chi^2_r$ is unity, as all other values are more extreme. 

\clearpage

\subsection{Residual plots of BIFT fits}

\begin{figure}
	\centering
	
	\includegraphics[width=\linewidth]{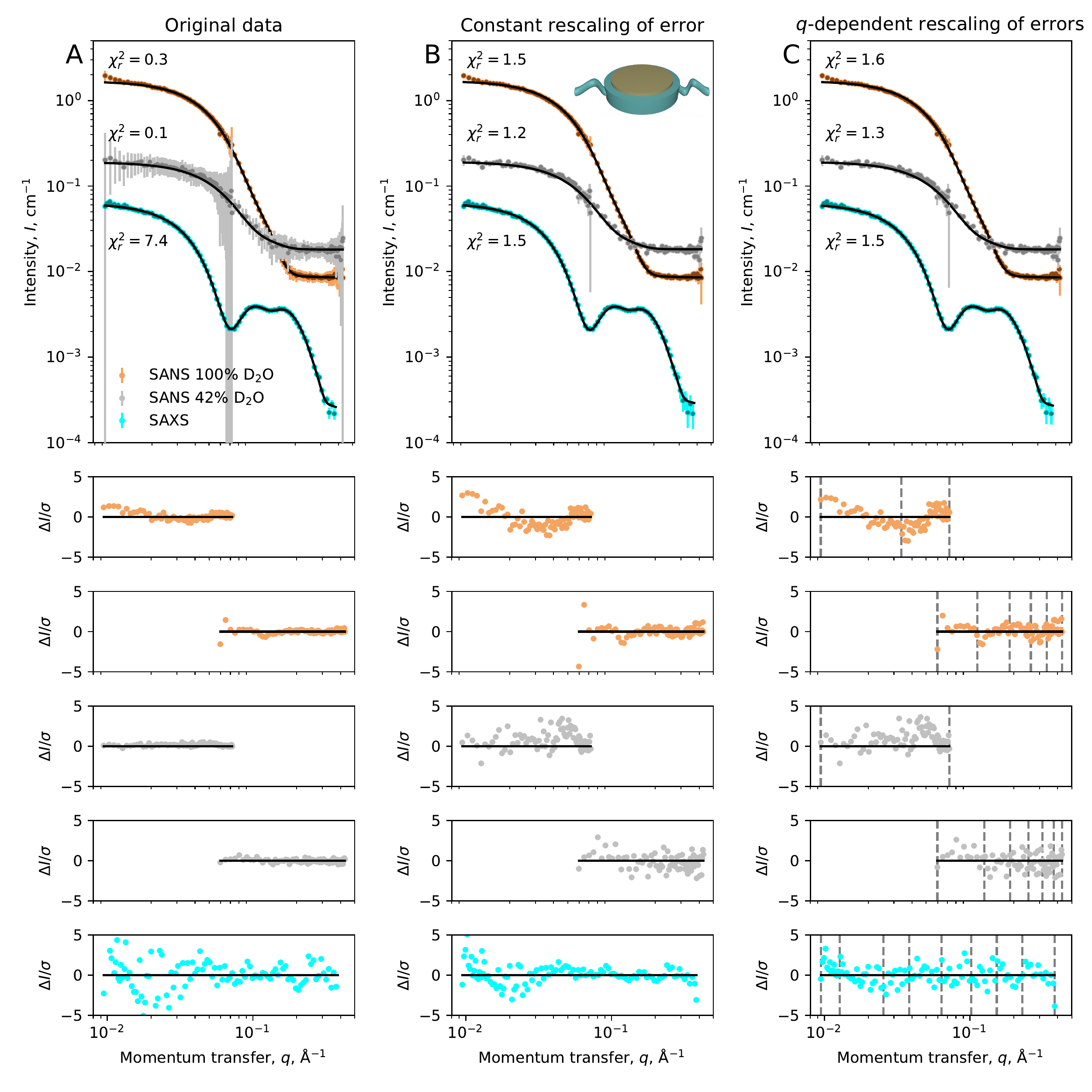}
	
	\caption{Plots and normalized residuals for the BIFT fit of the example data (Figure \ref{Figure:Example}), visualising $q$-dependency of the misestimated errors. The normalized residuals are expected to lie within a range of $-3$ and $3$, assuming only statistical noise. ()A) Original data and model fits. (B) Data and fits after rescaling experimental errors with a constant. Inset shows the geomerical model: a protein/lipid nanodisc. (C) Data and fits after rescaling errors in with factors varying along $q$. Shannon bins (see main text) with minimum $10$ datapoints in each bin are shown with vertical grey dashed lines in the residual plots.}
	
	\label{Figure:Residuals}
\end{figure}

Figure~\ref{Figure:Residuals} show the fits and residual plots of the data introduced in Figure~\ref{Figure:Example}. Of particular interest is the magnitude of the residuals in the various regions of the datasets, as this indicates the need for a $q$-dependent rescaling of the experimental errors, as done in panel C. 

\subsection{Simulations and tests of data with aggregation or hard-sphere potential tendencies}

\begin{figure}
	\centering
	
	\includegraphics[width=\linewidth]{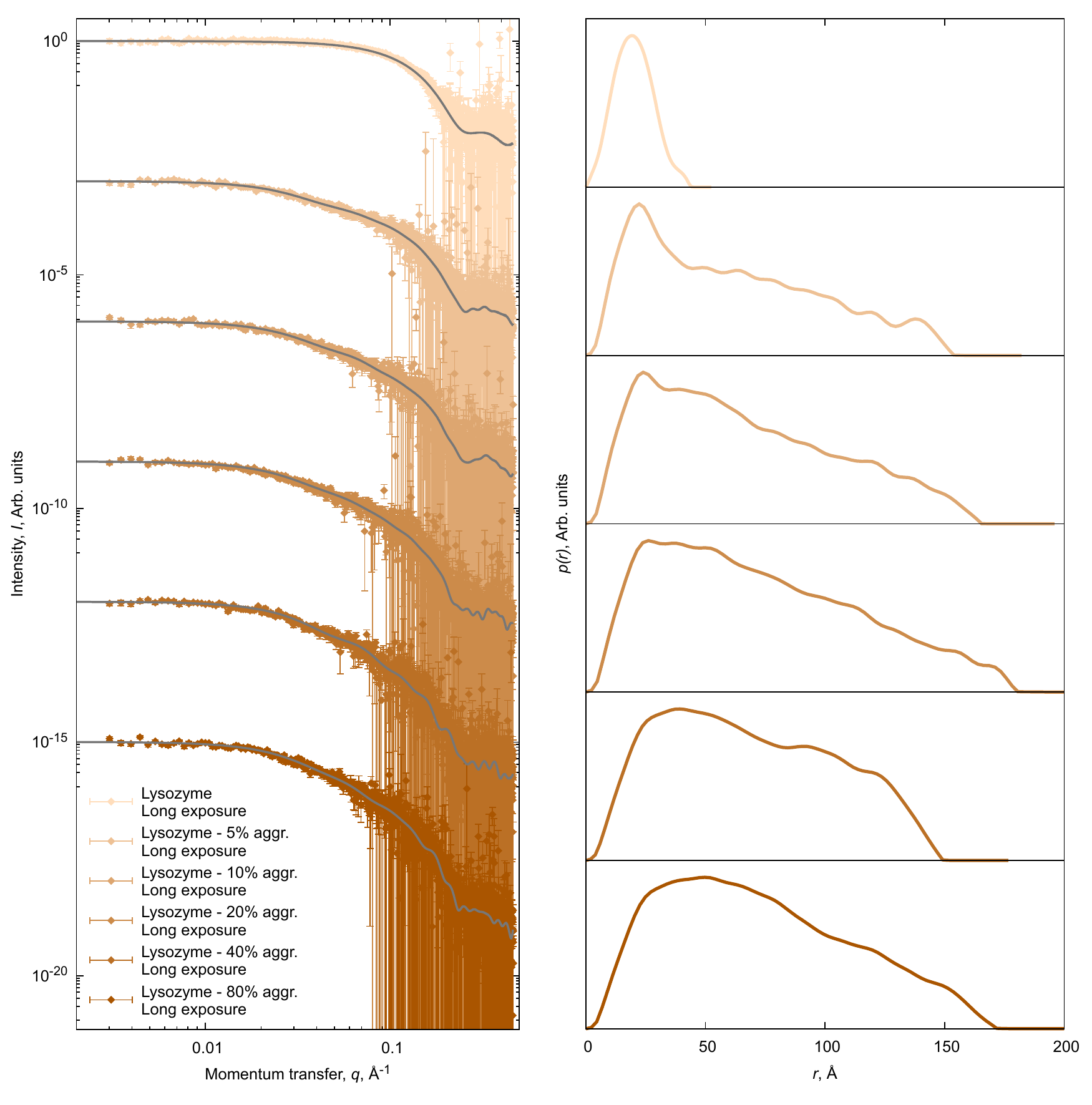}
	
	\caption{Simulated data and corresponding $p(r)$ distributions for lysozyme with aggregation. The grey lines are the BIFT fits.}
	
	\label{Figure:AggrData}
\end{figure}

\begin{figure}
	\centering
	
	\includegraphics[width=\linewidth]{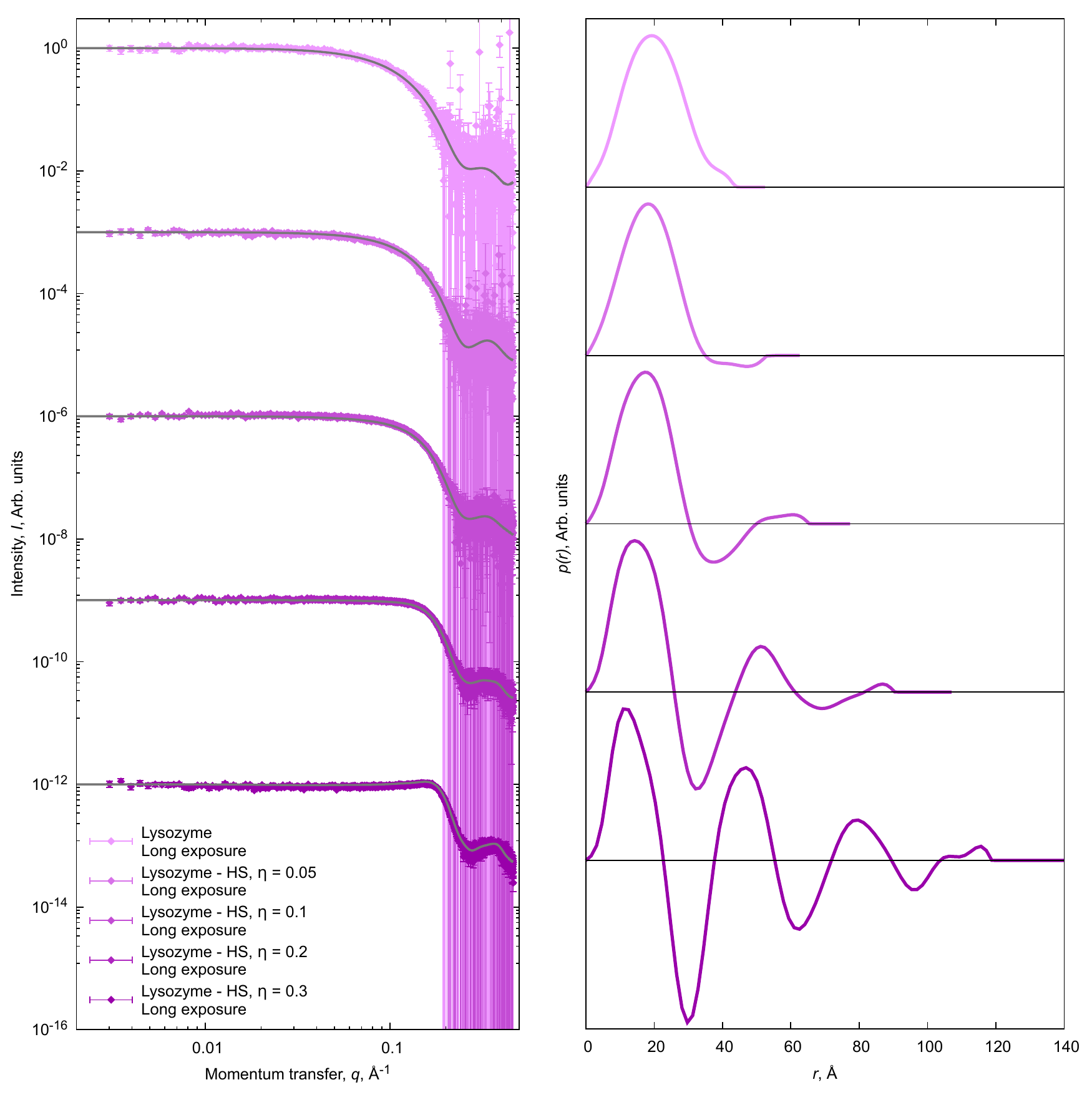}
	
	\caption{Simulated data and corresponding $p(r)$ distributions for lysozyme with a hard-sphere structure factor. The grey lines are the BIFT fits.}
	
	\label{Figure:HSData}
\end{figure}

\begin{figure}
	\centering
	
	\includegraphics[width=\linewidth]{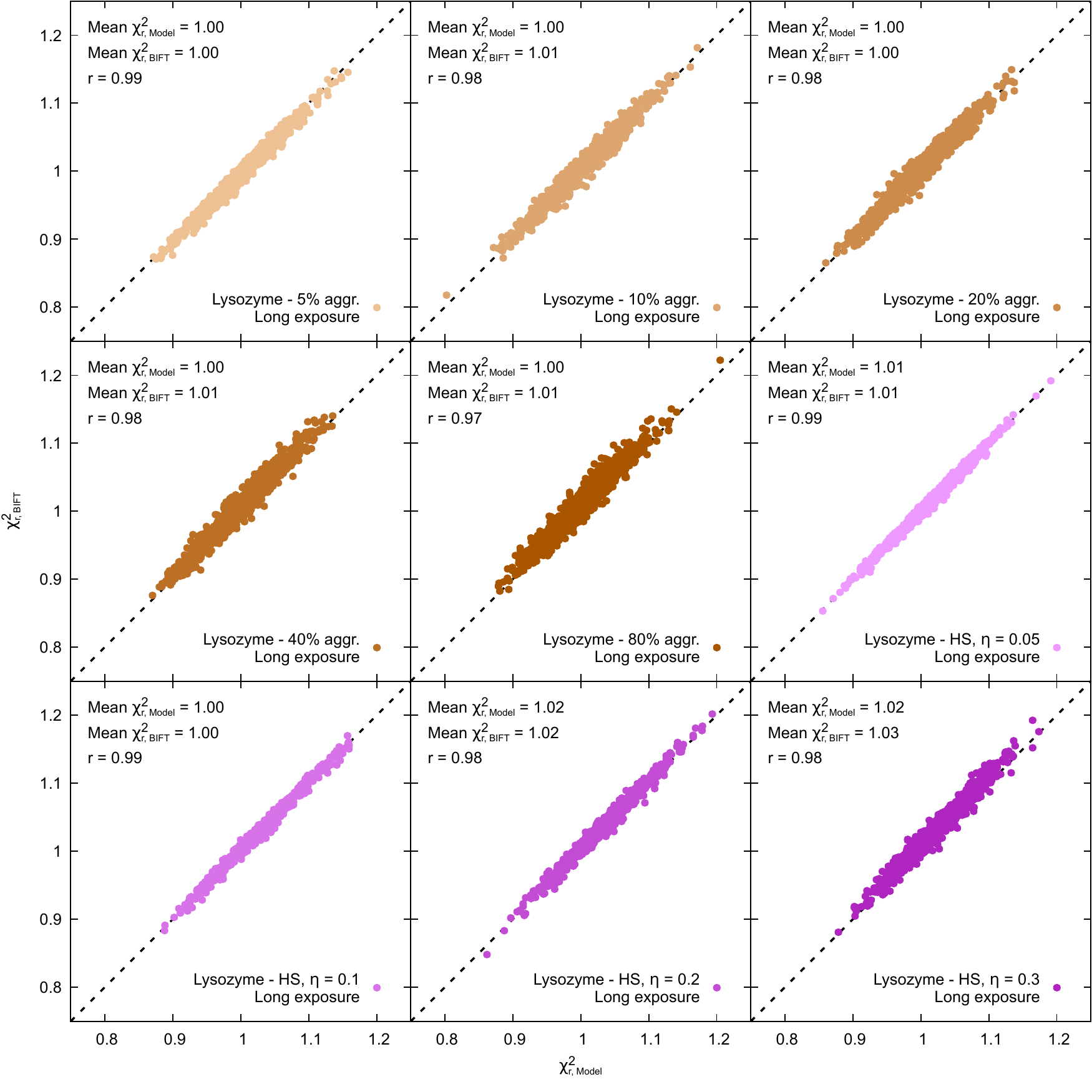}
	
	\caption{Correlation plots for the data simulated with either aggregation or a contribution from a hard-sphere potential. As previously, Pearson coefficients, $r$, and means are printed in the individual plots.}
	
	\label{Figure:SystematicCorrelation}
\end{figure}

Examples of the data simulated for this part of the study are shown in Figures~\ref{Figure:AggrData} and~\ref{Figure:HSData}. The plots establishing the correlation between the output of the BIFT algorithm and the simulated noise levels can be found in Figure~\ref{Figure:SystematicCorrelation}.

\end{document}